\documentclass[a4paper,10pt]{article} 

\usepackage{hyperref}
\usepackage{url}

\usepackage{authblk}

\usepackage{epsfig}
\usepackage{cite}
\usepackage{amsmath}
\usepackage{theorem}
\usepackage{graphicx}
\usepackage{amssymb}
\usepackage{latexsym}

\usepackage{color}

\definecolor{DarkBlue}{rgb}{0.1,0.1,0.5}
\definecolor{Red}{rgb}{0.9,0.0,0.1}
\definecolor{Purple}{rgb}{0.5,0,0.8}
\definecolor{Green}{rgb}{0.0,0.5,0}
\definecolor{Grey}{rgb}{0.3,0.3,0.3}
\definecolor{Yellow}{rgb}{0.9,0.9,0}

\newcommand{\mask}{\mathrm{mask}}
\newcommand{\Var}{\mathrm{Var}}

\renewcommand{\vec}[1]{\mathbf{#1}}
\newcommand{\grvec}[1]{\boldsymbol{#1}}
\newcommand{\tildebold}[1]{\tilde{\mathbf{#1}}}

\title{High-dimensional cluster analysis with the\\ Masked EM Algorithm}

\author[1]{Shabnam N.~Kadir \thanks{s.kadir@ucl.ac.uk} }

\author[2]{Dan F. M. Goodman \thanks{dan\_goodman@meei.harvard.edu} }

\author[1]{Kenneth D. Harris \thanks{kenneth.harris@ucl.ac.uk}}

\affil[1]{Institute of Neurology, Department of Neuroscience, Physiology, and Pharmacology\\
University College London\\
21 University Street\\London WC1E 6DE. }
\affil[2]{Eaton-Peabody Laboratories\\
 Massachusetts Eye and Ear Infirmary\\
Department of Otology and Laryngology\\
 Harvard Medical School.}

\begin{document}
\maketitle

\begin{abstract}
Cluster analysis faces two problems in high dimensions: first, the ``curse of dimensionality'' that can lead to overfitting and poor generalization performance; and second, the sheer time taken for conventional algorithms to process large amounts of high-dimensional data. In many applications, only a small subset of features provide information about the cluster membership of any one data point, however this informative feature subset may not be the same for all data points. Here we introduce a ``Masked EM'' algorithm for fitting mixture of Gaussians models in such cases. We show that the algorithm performs close to optimally on simulated Gaussian data, and in an application of ``spike sorting'' of high channel-count neuronal recordings.

\end{abstract}

\section{Introduction}

Cluster analysis is a widely used technique for unsupervised classification of data. A popular method for clustering is by fitting a mixture of Gaussians, often achieved using the Expectation-Maximization (EM) algorithm \cite{dempster1977maximum} and variants thereof \cite{fraley2002model}. In high dimensions however, this method faces two challenges \cite{bouveyron2012model}: first, the ``curse of dimensionality'' leads to poor classification, particularly in the presence of a large number of uninformative features; second, for large and high-dimensional data sets, the computational cost of many algorithms can be impractical. This is particularly the case where covariance matrices must be estimated, leading to computations of order $\mathcal{O}(p^2)$, where $p$ is the number of features; furthermore even a cost of $\mathcal{O}(p)$ can render a clustering method impractical for applications in which large high-dimensional data sets must be analyzed on a daily basis. In many cases however, the dimensionality problem is solvable at least in principle, as the features sufficient for classification of any particular data point are a small subset of the total available. 

A number of approaches have been suggested for the problem of high-dimensional cluster analysis. One approach consists of modifying the generative model underlying the mixture of Gaussians fit to enforce low-dimensional models. For example the Mixture of Factor Analyzers \cite{ghahramani1996algorithm,mclachlan2003modelling}  models the covariance matrix of each cluster as a low rank matrix added to a fixed diagonal matrix forming an approximate model of observation noise. This approach can reduce the number of parameters for each cluster from $\mathcal{O}(p^2)$ to $\mathcal{O}(p)$, and may thus provide a substantial improvement in both computational cost and performance. An alternative approach - which offers the opportunity to reduce the both the number of parameters and computational cost substantially below $\mathcal{O}(p)$ - is feature selection, whereby a small subset of informative features are selected and other non-informative features are discarded \cite{raftery2006variable}. A limitation of global feature selection methods however, is that they cannot deal with the case where different data points are defined by different sets of features. One proposed solution to this consists of assigning each cluster a unique distribution of weights over all features, which has been applied to the case of hierarchical clustering \cite{friedman2004clustering}.

The algorithm described below was developed to solve the problems of high-dimensional cluster analysis for a particular application: ``spike sorting'' of neurophysiological recordings using newly developed high-count silicon microelectrodes \cite{einevoll2012towards}. Spike sorting is the problem of identifying the firing times of neurons in the living brain, from electric field signatures recorded using multisite microfabricated neural electrodes \cite{lewicki1998review}. In a typical experiment, this will involve clustering millions of data points each of which reflects a single action potential waveform that could have been produced by one of many neurons. Historically, neural probes have had only a small number of channels (usually 4), typically resulting in feature vectors of 12 dimensions which required sorting into 10-15 clusters. Analysis of ``ground truth'' shows that the data is quite well approximated by a mixture of Gaussians with different covariance matrices between clusters \cite{einevoll2012towards}. Accordingly, in this low-dimensional case, traditional EM-derived algorithms work close to optimally \cite{harris2000accuracy}, although specialized rapid implementation software is required to cluster the millions of spikes recorded on a daily basis \cite{KK}. More recent neural probes, however, contain tens to hundreds of channels, spread over large spatial volumes, and probes with thousands are under development. Because different neurons have different spatial locations relative to the electrode array, each action potential is detected on only a small fraction of the total number of channels, but the subset differs between neurons, ruling out a simple global feature selection approach. Furthermore, because spikes produced by simultaneous firing of neurons at different locations must be clustered independently, most features for any one data point are not simply noise, but must be regarded as missing data. Finally, due to the large volumes of data produced by these methods, we require a solution that is capable of clustering millions of data points in reasonably short running time.

The Masked EM algorithm we introduce here works in two stages. In the first stage, a ``mask vector'' is computed for each data point via a heuristic algorithm, encoding a weighting of each feature for even data point. This stage may take advantage of domain-specific knowledge, such as the topological constraint that action potentials occupy a spatially contiguous set of recording channels. In the case that the majority of masks can be set to zero, the number of parameters per cluster and running time can both be substantially below $\mathcal{O}(p)$. We note that the masks are assigned to data points, rather than clusters, and need only be computed once at the start of the algorithm. The second stage consists of cluster analysis. This is implemented a mixture of Gaussians EM algorithm, but with every data point replaced by a virtual mixture of the original feature value, and the fixed subthreshold noise distribution weighted by the masks. The use of this virtual mixture distribution avoids the splitting of clusters due to arbitrary threshold crossings. At no point is it required to generate samples from the virtual distribution, as expectations over it can be computed analytically.

\section{The Masked EM Algorithm}
\label{MKK}

\begin{table}[t]
\caption{Mathematical Notation}
\label{notation_table}
\begin{center}
\begin{tabular}{c|c}\hline
\multicolumn{2}{c}{Notation}\\\hline
Dimensions (number of features) & $p$\\
Data (point $n$, feature $i$) & $x_{n,i}$\\
Masks & $m_{n,i}\in[0,1]$\\
Cluster label &$k$\\
Total number of clusters & $K$\\
Mixture weight, cluster mean, covariance&$w_k,\grvec{\mu}_k,\grvec{\Sigma}_k$\\
Probability density function of multivariate Gaussian distribution&$\phi\left(\vec{x}|\grvec{\mu}_k, \grvec{\Sigma}_k\right)$\\
Total number of data points & $N$\\
Number of points for which feature $i$ is masked &$N_i^{\mask}$\\
Noise mean for feature $i$ & $\nu_i $\\
Noise variance for feature $i$&$\sigma_i^2$\\
Virtual features (random variable) & $\tilde x_{n,i}$ \\
Mean of virtual feature & $y_{n,i}=\mathbb{E}[\tilde x_{n,i}]$\\
Variance of virtual feature & $\eta_{n,i}:=\Var(\tilde x_{n,i})$\\
Logarithm of the responsibilities&$\pi_{nk}$\\
Set of data points assigned to cluster $k$ &$\mathcal{C}_k$\\
Subset of $\mathcal{C}_k$ for which feature $i$ is fully masked &$\mathcal{M}_{ik}$\\
\hline
\end{tabular}
\end{center}
\end{table}

\subsection{Stage 1: mask generation}

The first stage of the algorithm consists of computing a set of ``mask vectors'' indicating which features should be used to classify which data points. Specifically, the outcome of this algorithm is a set of vectors $\vec{m}_n$ with components, $m_{n,i}\in[0,1]$, with a value of $1$ indicating that feature $i$ is to be used in classifying data point $\vec{x}_n$, a value of $0$ indicating it is to be ignored, and intermediate values corresponding to partial weighting. We refer to features being used for classification as ``unmasked'', and features being ignored as ``masked'' (i.e. concealed). The use of masks provides two major advantages over a standard mixture of Gaussians classification: first, it overcomes the curse of dimensionality, because assignment of points to classes is no longer overwhelmed by the noise on the large number of masked channels; second, it allows the algorithm to run in time proportional to $\mathcal{O}(\text{unmasked features}^2)$ rather than $\mathcal{O}(\mbox{features}^2)$. Because a small number of features may be unmasked for each data point, this can allow computational costs substantially below $\mathcal{O}(p)$ The way masks are chosen can depend on the application domain and would typically follow a heuristic method. A simple approach that can work in general is a standard deviation threshold: 
\begin{eqnarray}
m_{n,i} = \begin{cases} 0 & |x_{n,i}|<\alpha SD_i\\
1 & |x_{n,i}|>\beta SD_i \\
\frac{x-\alpha SD_i}{\beta SD_i-\alpha SD_i} &\alpha SD_i< |x_{n,i}|<\beta SD_i \\
\end{cases}
\label{threshold}
\end{eqnarray}
In the case of spike sorting, a slightly more complex procedure is used to derive the masks, which takes advantage of the topological constraint that spikes must be distributed across continuous groups of recording channels. The software for this, SpikeDetekt, can be downloaded from  \url{https://github.com/klusta-team/spikedetekt} \cite{SD}.  In practice, we have found that the choice of masking algorithm is not critical, provided the majority of non-informative features have a mask of $0$, and that features that are clearly suprathreshold are given a mask of 1.

Once the masks have been computed, an additional set of quantities is pre-computed before the main EM loop starts. Specifically, the subthreshold noise mean for feature $i$, $\nu_i$ is obtained by taking the mean of feature $i$ whenever that particular feature is masked, i.e. $m_{n,i}= 0$, and analogously, the noise variance for feature $i$, $\sigma_i^2$:
 $$\nu_i := \frac{1}{N_i^{\mask}}\sum_{n:m_{n,i} = 0} x_{n,i},\quad \sigma_i^2:= \frac{1}{N_i^{\mask}}\sum_{n:m_{n,i} = 0}(x_{n,i}-\nu_i)^2,$$

where $N_i^{\mask} =|\{n: m_{n,i} = 0\}| $.

\subsection{Stage 2: clustering}

The second stage consists of a maximum-likelihood mixture of Gaussians fit, with both the E and M steps modified by replacing each data point $\vec{x}_n$ with a virtual ensemble of points $\tildebold{x}_n$ distributed as

\begin{eqnarray}
\tilde x_{n,i} = \begin{cases} x_{n,i} &\mbox{prob  }\ m_{n,i}\\
                          N(\nu_i,\sigma_i^2) &\mbox{prob}\  1-m_{n,i},\\
                       \end{cases}
\end{eqnarray}

where  $m_{n,i}\in[0,1]$ is the mask vector component associated to $x_{n,i}$ for the $n$th spike. Intuitively, any feature below a noise threshold is replaced by a `virtual ensemble' of the entire noise distribution. The noise on each feature will be modeled as independent univariate Gaussian distributions $N(\nu_i,\sigma_i^2)$ for each $i$, which we shall refer to as the noise distribution for feature $i$. This is, of course, a simplification, as the noise may be correlated. However for tractability, ease of implementation and as we shall later show, efficacy, this approximation suffices.

The algorithm maximizes the expectation of the usual log likelihood over the virtual distribution:

\begin{eqnarray}
L\left(w_k,\grvec{\mu}_k,\grvec{\Sigma}_k\right)&:=&\sum_{n=1}^{N} \mathbb{E}_{\tildebold x}\left[ \log\left(\sum_{k=1}^{K} w_k\frac{\exp\left(-\frac{1}{2}(\tildebold x_n-\grvec{\mu}_k)^T \grvec{\Sigma}_k^{-1}(\tildebold x_n-\grvec{\mu}_k)\right)}{(2\pi)^{d/2}\|\grvec{\Sigma}_k\|^{1/2}}\right)\right].\nonumber
\end{eqnarray}

The masked EM algorithm therefore acts as if it were passed an ensemble of points, with each data point replaced by an infinite sample, corresponding to different possibilities for the noisy masked variables. This solves the curse of dimensionality by ``disenfranchising'' each data point's masked features, disregarding the value that was actually measured, and replacing it by a virtual ensemble that is the same in all cases, and thus does not contribute to cluster assignment.

Before running the EM algorithm, the following quantities are also computed, which will greatly speed up computation of the modified M and E-steps:
$$y_{n,i} := \mathbb{E}[\tilde x_{n,i}] = m_{n,i}x_{n,i} + (1-m_{n,i})\nu_i, $$
$$z_{n,i} := \mathbb{E}[(\tilde x_{n,i})^2] = m_{n,i}(x_{n,i})^2 + (1-m_{n,i})(\nu_i^2 + \sigma_i^2), $$
$$\eta_{n,i} := \Var [ \tilde x_{n,i}] = z_{n,i} - (y_{n,i})^2.$$

\subsection{M-step}
For the M-step, replacing $\vec{x}$ with the virtual ensemble $\tildebold x$ requires computing the expectation with respect to $\tilde x_{n,i}$ of the mean and the covariance of each cluster. For simplicity, we henceforth focus on a ``hard'' EM algorithm in which each data point $\vec{x}_n$is fully assigned to a single cluster, although a ``soft'' version can be easily derived. We denote by $\mathcal{C}_k$  the set of data point indices assigned to the cluster with index $k$. It is straightforward to show that:

\begin{equation}
(\grvec{\mu}_k)_i  = \frac{1}{|\mathcal{C}_k|}\sum_{n \in C} y_{n,i},\ \  (\grvec{\Sigma}_k)_{ij}= \mathbb{E}[(\tildebold{ \grvec{\Sigma}}_k)_{ij}] = \frac{1}{|\mathcal{C}_k|}\sum_{n \in C} \left( y_{n,i}-(\grvec{\mu}_k)_i)(y_{n,j}-(\grvec{\mu}_k)_j) + \eta_{n,i}\delta_{i,j} \right)
\end{equation}
 

Note that this is the same formula as the classical M-step using, but with $x_{n,i}$ replaced by the expected value $y_{n,i}$ of the virtual distribution $\tilde x$, plus a correction term to $\Sigma_{i,j}$ corresponding to the covariance matrix $\eta_{n,i}$ of $\tilde x$. Computation of these quantities can be carried out very efficiently as we can decompose $(\grvec{\mu}_k)_i $ and  $(\grvec{\Sigma}_k)_{ij} $ as follows:

\begin{equation}
(\grvec{\mu}_k)_i = \frac{1}{|\mathcal{C}_k|}\left(\sum_{n \in \mathcal{C}_k\setminus \mathcal{M}_{k,i}} y_{n,i}+|\mathcal{M}_{k,i}|\nu_i\right) ,
\end{equation}

{\small
\begin{eqnarray*}
(\grvec{\Sigma}_k)_{ij} &=& \frac{1}{|\mathcal{C}_k|} \sum_{n \in \mathcal{C}_k\setminus (\mathcal{M}_{k,i} \cap\ \mathcal{M}_{k,j}) } (y_{n,i}-(\grvec{\mu}_k)_i )(y_{n,j}-(\grvec{\mu}_k)_i )\\
&& +\frac{ |\mathcal{M}_{k,i}\cap \mathcal{M}_{k,j} |}{|\mathcal{C}_k|} (\nu_{i}-(\grvec{\mu}_k)_i)(\nu_{j}-(\grvec{\mu}_k)_j)  +\frac{1}{|\mathcal{C}_k|}\left(\sum_{n \in \mathcal{C}_k\setminus \mathcal{M}_{k,i} }\eta_{n,i} + |\mathcal{M}_{k,i}| \sigma_{i}^2\right) \delta_{i,j},\\
\end{eqnarray*}
}
where $\mathcal{M}_{k,i} = \{n\in \mathcal{C}_{k} | m_{n,i} = 0 \}\subseteq \mathcal{C}_{k}$ denotes the set of points within cluster $k$ for which feature $i$ is fully masked. Note that if all data points in a cluster have feature $i$ masked, then $(\grvec{\mu}_k)_i = \nu_i$, the noise mean, and $(\grvec{\Sigma}_k)_{ii} = \sigma_{i}^2$, the noise variance.

\subsection{E-step}
For the E-step the log of the responsibilities $\pi_{nk}$ is computed as its expected value over the virtual distribution $\tilde x$. Thus, the algorithm acts as if each data point is replaced by an infinite ensemble of points drawn from the distribution of $\tilde x$, which must all be assigned the same cluster label. Explicitly,

{\small
\begin{eqnarray}
\pi_{nk} = \mathbb{E}_{\tilde x} \left[ -\frac{d}{2} \log 2 \pi -\frac{1}{2}\log \det \grvec{\Sigma}_k - \frac{1}{2}(\tildebold{x}_n-\grvec{\mu}_k)^T(\grvec{\Sigma}_k)^{-1}(\tildebold{x}_n-\grvec{\mu}_k)\right] .
\label{logresp}
\end{eqnarray}
}


The final term of  \ref{logresp} corresponds to the expectation of the Mahalanobis distance of $\tilde x_{n,i}$ from cluster $k$; it can be shown that 
{\small
\begin{eqnarray*}
\pi_{nk} 
&=&  -\frac{d}{2} \log 2 \pi -\frac{1}{2}\log \det \grvec{\Sigma}_k - 
\frac{1}{2}(\vec{y}_{n}-(\grvec{\mu}_k))^T(\grvec{\Sigma}_k)^{-1}(\vec{y}_{n}-(\grvec{\mu}_k))-\frac{1}{2}\left(\sum_{i} \eta_{n,i}(\grvec{\Sigma}_k)_{ii}^{-1}\right)
\end{eqnarray*}
}

This leads to the original E-step for the EM algorithm, but with $y_{n,i}$ substituted for $x_{n,i}$, plus a diagonal correction term $-\frac{1}{2}\sum_{i} \eta_{n,i}(\grvec{\Sigma}_k)_{ii}^{-1}$.

\subsection{Penalties}
In order to automatically determine the number of clusters in a mixture of Gaussians requires a penalty function, that penalises over-fitting by discouraging models with a large number of parameters. Commonly used penalization methods include the Akaike Information Criterion (AIC) \cite{stibor:Akaike:1974} and Bayes Information Criterion (BIC) \cite{schwarz1978estimating}:
$$\mbox{AIC} = 2\kappa - 2\ln(L), \quad \mbox{BIC} = \kappa\ln(N) - 2\ln(L),$$ where $\kappa$ is the number of free parameters in the statistical model, and $L$ is the maximum of the likelihood for the estimated model and $N$ is the number of data points.

For the a classical mixture of Gaussians fit, the number of parameters $\kappa$ is equal to $K\left(\frac{p(p+1)}{2} + p + 1\right)-1$
where $K$ is the number of clusters and $p$ is number of features. The elements of the first term in $\kappa$ correspond to the number of free parameters in a $p\times p$  covariance matrix, a $p$-dimensional mean vector, and a single weight for each cluster. Finally, $1$ is substracted from the total because of the constraint that the weights must sum to $1$ for a mixture model.

For masked data the estimation of the number of parameters in the model is more subtle. Because masked features are replaced by a fixed distribution that does not vary between data points, the effective degrees of freedom per cluster is much smaller than $\frac{p(p+1)}{2} + p + 1$. We therefore define a cluster penalty for each cluster $C$ that depends only on the average number of unmasked features corresponding to that cluster. Specifically, let $r_n:= \sum_{j=1}^{p} m_{n,j}$ be the effective number of unmasked features for data point $n$ (i.e. sum of the weights over the features).
Define $F(r) := \frac{r(r+1)}{2} + r + 1,$
where the three terms correspond to the number of free parameters in an $r\times r$ covariance matrix, mean vector of length $r$ and a mixture weight, respectively.

Our estimate of the effective number of parameters is thus:
\begin{eqnarray}
 \hat{\kappa}= \sum_{k=1}^{K}\left(\frac{1}{|\mathcal{C}_k|}\sum_{n=1}^{|\mathcal{C}_k|}F(r_n)\right) - 1
\end{eqnarray}

In practice, we have found that substituting this formula for the effective number of parameters into AIC provides good performance (see below).

\subsection{Implementation}

The algorithm was implemented in custom C++ code, based on previously released open-source software for fast mixture of Gaussians fitting termed ``KlustaKwik'' \cite{KK}. Because we require the algorithm to run very fast on millions of high-dimensional data points, several approximations are made to give faster running time without significant impact on performance. These include not only hard classification but also a heuristic that eliminates the great majority of E-step iterations;  a split-and-merge feature that changes the number of clusters dyamically if this increases the penalized likelihood; and an additional uniform distributed mixture component to catch outliers. The software can be downloaded from \url{https://github.com/klusta-team/klustakwik} \cite{MKK}.

\section{Evaluation}

\subsection{Mixture of Gaussians}
\begin{figure}[htbp]
\begin{center}
\includegraphics[width=1\linewidth]{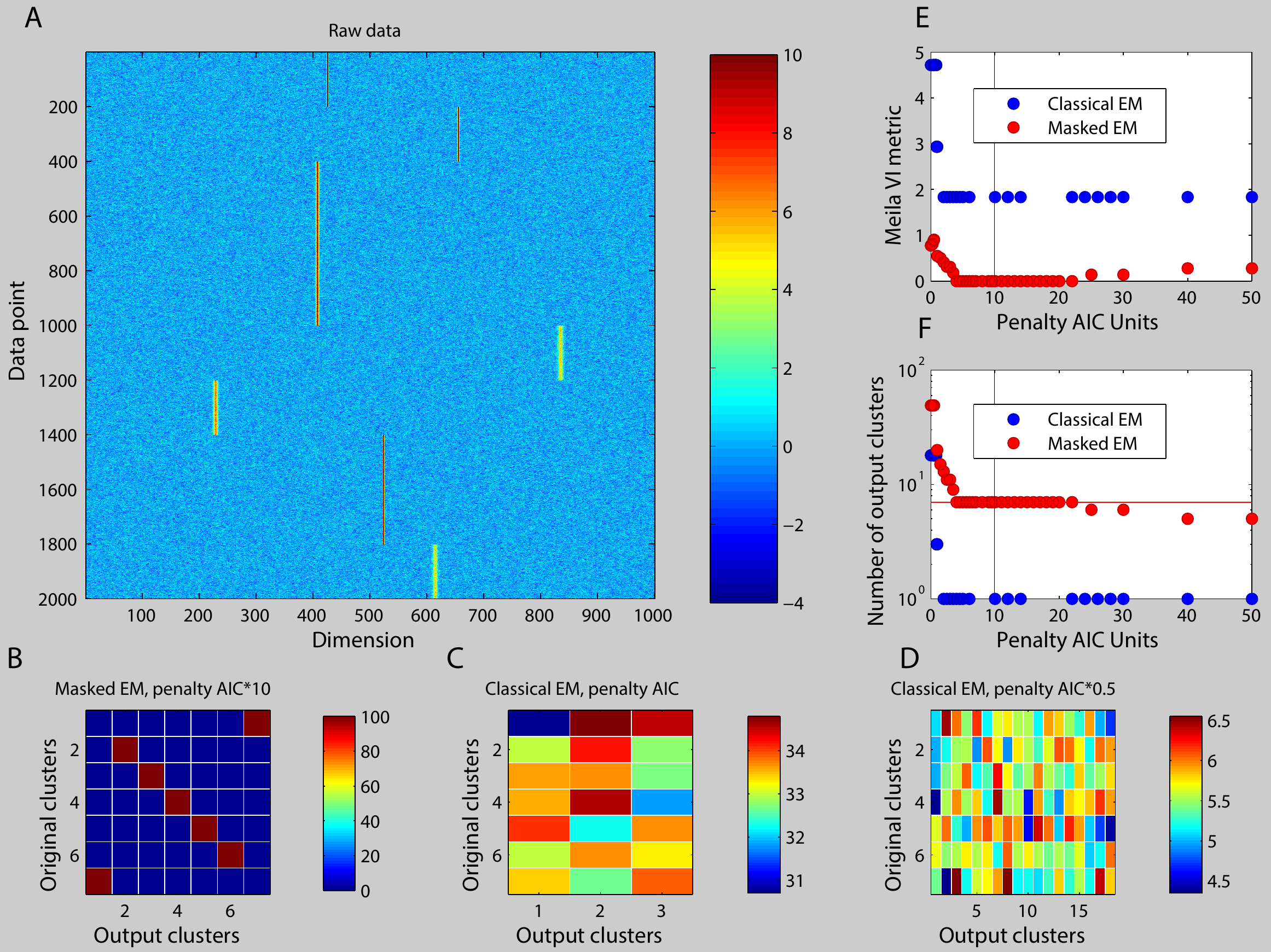}
\end{center}
\caption{Simulated Data. {\bf A}: Subsampled raw data, {\bf B}: Confusion matrix in percentages for Masked EM for an BIC ($10*\mbox{AIC}\approx\log 20000*\mbox{AIC}$) penalty, {\bf C}: Confusion matrix in percentages for Classical EM for an AIC penalty, {\bf D}:
Confusion matrix in percentages for Classical EM for a penalty of $0.5\ast\mbox{AIC}$, {\bf E}: VI metric measure of performance of both algorithms using various values for penalty, where the black vertical line indicates BIC, {\bf F}: The number of clusters obtained for various values of penalty, where the black vertical line indicates BIC.  }
\label{simulated}
\end{figure}

We first demonstrate the efficacy of the Masked EM algorithm using a simple data set synthesized from a high-dimensional mixture of Gaussians. The data set consisted of $20000$ points in  $1000$ dimensions, drawn from $7$ separate clusters. The means were chosen by centering probability density functions of gamma distributions on certain chosen features. All covariance matrices were identical, namely a Toeplitz matrix with all the diagonal entries $1$ and off-diagonal entries that decayed exponentially with distance from the diagonal. Figure \ref{simulated}A shows this data set in pseudocolor format.

Figure \ref{simulated}B shows a confusion matrix generated by the Masked EM algorithm on this data, with the modified $\mbox{AIC}*10\approx \mbox{AIC}*\log 20000 \equiv \mbox{BIC}$  penalty and masks defined by equation \ref{threshold} , indicating perfect performance. By contrast, Figure \ref{simulated}C shows the result of classical EM, in which many clusters have been erroneously merged. We are showing the results for AIC penalty since using a BIC penalty yielded only a single cluster. To verify that this is not simply due to an inappropriate choice of cluster numbers, we reran with the penalty term linearly scaled by various factors. Figure \ref{simulated}D shows the results of a penalty $0.5\ast\mbox{AIC}$ that gave a larger number of clusters; even in this case, however, classification performance was poor. To systematically evaluate the effectiveness of both algorithms, we measured performance using the Variation of Information  (VI) metric \cite{meilua2003comparing}, for which a value of $0$ indicates a perfect clustering. Both algorithms were tested for a variety of different penalties measured in multiples of AIC (Figure \ref{simulated}E,F). Whereas the Masked EM algorithm was able to achieve a perfect clustering for a large range of penalties around BIC, the classical EM algorithm was only able to produce a poorer value of $1.83$ (corresponding to the poor result of merging all the points into a single cluster).

\subsection{Spike sorting}

To test the performance of the Masked EM algorithm on our target application of high-channel-count spike sorting requires a ground truth data set. Previous work established the performance of the classical EM algorithm for low-channel spike sorting with ground truth obtained by simultaneous recordings of a neuron using not only the extracellular array, but also an intracellular using a glass pipette that unequivocally determined firing times \cite{harris2000accuracy}. Because such dual recordings are not yet available for high-count elecrodes, we created a simulated ground truth we term ``hybrid datasets''. In this approach, a mean spike waveform of a single neuron taken from one recording (the ``donor'') is digitally added onto  a second recording (the ``acceptor'') made with the same electrode in a different brain. Because the ``hybrid spikes'' are linearly added to the acceptor traces, this simulates the linear addition of neuronal electric fields, and recreates many of the challenges of spike sorting such as overlapping spikes \cite{harris2000accuracy}. The hybrid datasets we consider were constructed  from recordings in rat cortex kindly provided by Mariano Belluscio and Buzs{\'a}ki, Gy{\"o}rgy, using a 32-channel probe with a zig-zag arrangement of electrodes and minimum $20\mu$m spacing between neighbouring contacts.
Three principal components are taken from each channel resulting $96$-dimensional feature vectors.

\begin{figure}[htbp]
\begin{center}
\includegraphics[width=0.8\linewidth]{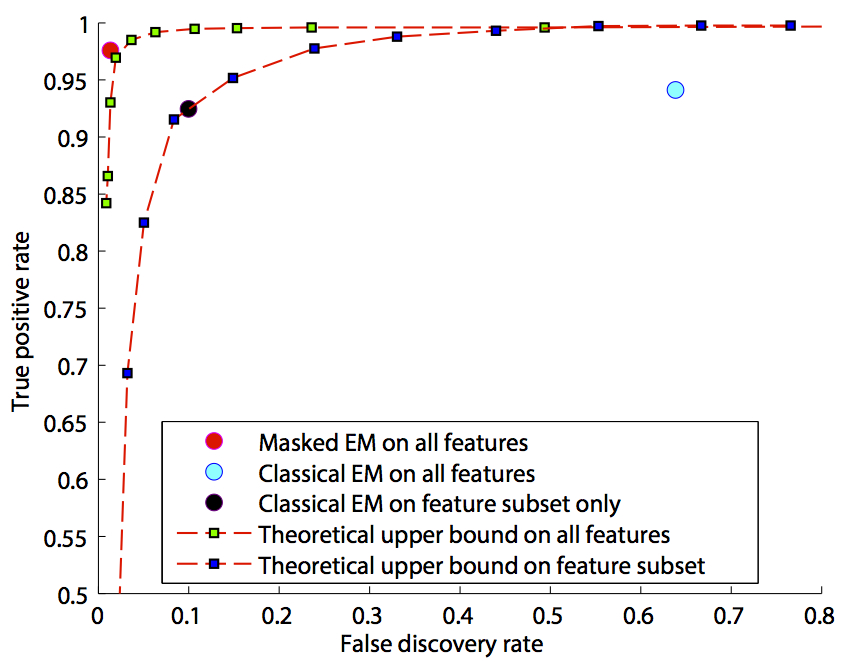}
\end{center}
\caption{ Performance of the masked and classical EM algorithms in a spike sorting application}
\label{SVMvsKKMKK}
\end{figure}

To evaluate the performance of the algorithm, we first identified the cluster with the highest number of true positive spikes, and used it to compute a false discovery rate,  $\frac{FP}{FP+TP}$, and a true positive rate $\frac{TP}{FN+TP}$, where FP denotes the number of false positive, TP, the number of true positive, FN, the number of false negative spikes. This performance was compared against a theoretical upper bound obtained by supervised learning. The upper bound was obtained by using a support vector machine (SVM) \cite{cortes1995support} with inhomogeneous quadric kernel trained using the ground truth data, with performance evaluated by 2-fold cross-validation. By varying the weighting of false positive and false negative errors, we obtained a receiver operating characteristic (ROC)-curve. The value of the margin parameter $C$ was chosen as that optimizing cross-validation performance.

Figure \ref{SVMvsKKMKK} shows the performance of the Masked EM algorithm and classical EM algorithm on the hybrid data set, set against the theoretical optimum revealed by the SVM. While the Masked EM algorithm performs at close to the upper bound, the classical EM  algorithm is much poorer. To verify that this poorer performance indeed resulted from a curse of dimensionality, we re-ran the classical EM algorithm on only the subset of features that were unmasked for the hybrid spike ($9$ out of $96$ features). As expected, the upper-bound performance is somewhat poorer in this case, but the classical EM algorithm now performs close to the theoretical upper bound. This indicates that the classical algorithm fails in high dimensional settings, whereas the Masked EM algorithm performs well. 

\section{Discussion and Conclusions}

We have introduced a method for high-dimensional clustering, applicable to the case where a small subset of a large number of potential features is informative for any data point. Unlike global feature selection methods, both the number and the precise set of unmasked features can vary between different data points. Both the number of free parameters and computational cost scale with the number of unmasked features per data point, rather than the total number of features. This approach was found to give good performance on simulated high-dimensional data, and in our target application of neurophysiological spike sorting for large electrode arrays.

A potential caveat of allowing different features to define different clusters is the danger of artificial cluster splitting. If a hard threshold were simply used to decide whether a particular feature should be used for a particular cluster or data point, this could lead to a single cluster being erroneously split in two, according to whether or not the threshold was exceeded by noisy data. The Masked EM algorithm overcomes this problem with two approaches. First, because the masks are not binary but real-valued, crossing a threshold such as that in equation \ref{threshold} leads to smooth rather than discontinuous changes in responsibilities; second, because masked features are replaced by a virtual distribution of empirically measured subthreshold data, the assignment of points with masked features is close to that expected if the original subthreshold features had not been masked. With these approaches in place, we found that erroneous cluster splitting was not a problem in simulation or in our target application.

In the present study, we have applied the masking strategy to a single application of unsupervised classification using a hard EM algorithm for mixture of Gaussians fitting. However the same approach may apply more generally whenever only a subset of features are informative for any data point, and when the expectation of required quantities over the modeled subthreshold distribution can be analytically computed. Other domains in which this approach may work therefore include not only cluster analysis with soft EM algorithms or different probabilistic models, but also model-based supervised classification.

\bibliography{arxiv2013cluster}{}
\bibliographystyle{unsrt}

\end{document}